# Capture cross sections of $^{15}$N(n, γ)$^{16}$N at astrophysical energies

FAN Guang-wei[1], MA Jun-bing[2], SHENG Zong-qiang[1], TIAN Feng[3], WANG Jun[1], ZHANG Chao[1]

[1] School of Chemical Engineering, Anhui University of Science and Technology, Huainan, 232001, China
[2] Institute of Modern Physics, Chinese Academy of Sciences, Lanzhou, 730000, China
[3] Shanghai Institute of Applied Physics, Chinese Academy of Sciences, Shanghai, 201800, China

**Abstract:** We have reanalyzed reaction cross sections of $^{16}$N on $^{12}$C target. The nucleon density distribution of $^{16}$N, especially surface density distribution, was extracted using the modified Glauber model. On the basis of dilute surface densities, the discussion of $^{15}$N(n, γ)$^{16}$N reaction was performed within the framework of the direct capture reaction mechanism. The calculations agreed quite well with the experimental data.

## 1 Introduction

Nuclear reactions at low energies play a crucial role in nuclear astrophysics. Often the reactions are quite difficult to measure in laboratories; the theoretical extrapolation is important in the studies [1-4]. The $^{15}$N(*n, γ*)$^{16}$N reaction is such a low-energy reaction, its cross section ($\sigma_c$) at astrophysical energies is an important input in the reaction network for the determination of heavier neutron-rich elements A > 16 in both inhomogeneous big bang and in red giant environments. Besides, it is a competition reaction to the reaction $^{15}$N(*α, γ*)$^{19}$F. $^{19}$F is a key element for evolutive studies in the asymptotic giant branch stars in which the $^{15}$N(*n, γ*)$^{16}$N reaction largely affect the abundance of $^{19}$F [5-7].

The experimental $\sigma_c$ of $^{15}$N(*n, γ*)$^{16}$N reaction are of considerable uncertainties. In 1996, Meissner *et al*. [8] measured the $\sigma_c$ at neutron energies of 25, 152, and 370 keV. The authors have performed direct capture calculations to interpret the measurements, however, a large gap appears between the calculations and the experimental data. Thus, it is necessary for the further theoretical studies. In the process of the $^{15}$N(*n, γ*)$^{16}$N reaction, a free neutron at the continuum state is captured by the $^{15}$N target and finally stays in a ground state of the compound nucleus $^{16}$N. The reaction is mostly determined by the spectroscopic factors, the nuclear structure properties of four low-lying states in $^{16}$N, and the effective interaction potential between the free neutron and the $^{15}$N target. As for the spectroscopic factors, Bohne *et al*. [9] and Bardayan *et al*. [10] measured the angular distributions of $^{15}$N(*d, p*)$^{16}$N reaction, Guo *et al*. [11] measured the angular distributions of $^{15}$N(*$^{7}$Li, $^{6}$Li*)$^{16}$N reaction. These groups obtained the spectroscopic factors, respectively. Although a recent experiment and analysis for $^{15}$N(*n, γ*)$^{16}$N by Guo *et al*. gives a more accurate calculation of this reaction in the energy regime of interest, it is still meaningful to explore the errors from the structure of $^{16}$N. Fan *et al*. [12] have proved that the low-lying states structure in $^{8}$Li and the interaction potential between $^{7}$Li and free neutron can well been explored by the reaction cross section ($\sigma_R$) of $^{8}$Li on stable targets. In this paper, we will analyze the errors of $^{15}$N(*n, γ*)$^{16}$N according to the low-lying state structure of $^{16}$N deduced from $\sigma_R$ of $^{16}$N on $^{12}$C target.

## 2 Theoretical mechanism

The Glauber model is a powerful tool to extract the nuclear surface structure by fitting the experimental $\sigma_R$. The model is based on the independent individual nucleon-nucleon (N-N) collisions in the overlap zone of the colliding nuclei, which account for a significant part of breakup effects. It successfully explains the observed $\sigma_R$ for various systems [13, 14]. The model will be employed in the article to deduce the surface structure of $^{16}$N and interaction potential of incident neutron on the $^{15}$N core. The Glauber model is a standard calculation, details can be found in a number of Refs. [15-17].

On the basis of the experimental spectroscopic factors, surface structure of $^{16}$N, and the interaction potential, the direct capture theory will be utilized to calculate the $\sigma_c$ of the reaction $^{15}$N$(n, \gamma)^{16}$N [18]. In the theory the neutron in a continuum state is captured by a target nucleus $^{15}$N that goes to a composite nucleus $^{16}$N via a transition with an $E1$ electric dipole. The $\sigma_c$ is given by

$$\sigma_{^{15}N \to ^{16}N}^{E1} = \frac{16\pi}{9\hbar} \kappa_\gamma^3 \left| Q_{^{15}N \to ^{16}N}^{E1} \right|, \tag{1}$$

where $\kappa_\gamma = \frac{E_\gamma}{\hbar c}$ is the wave number that corresponds to a $\gamma$-ray energy $E_\gamma$. $Q_{^{15}N \to ^{16}N}^{E1}$ transition matrix element given by

$$Q_{^{15}N \to ^{16}N}^{E1} = \langle \psi_c | O^{E1} | \psi_c \rangle, \tag{2}$$

where $O^{E1}$ stands for the electric dipole operator. The initial state wave function $\psi_c$ is the incoming neutron wave function, and the wave function $\psi_b$ represents the bound state of the composite nucleus $^{16}$N. The wave functions necessary in the direct theory will be obtained by solving the scattering and bound-state systems, respectively, for a given interaction potential. Thus, the essential ingredients are the potentials used to generate the wave functions $\psi_c$ and $\psi_b$ and the normalization given by the spectroscopic factor.

## 3 Nuclear structure of $^{16}$N

Ozawa et al. [19] and Fang et al. [20] have measured the $\sigma_R$ of $^{16}$N on $^{12}$C target, analyzed the experimental data with standard Glauber model, and obtained the density distribution of $^{16}$N. Unfortunately, there is a 10% - 20% underestimation in the standard Glauber model between the experimental $\sigma_R$ and the theoretical calculations at intermediate energies [21]. Although the density distribution has been extracted, it is necessary to reanalyze the experimental data with modified Glauber model [22]. The Glauber model requires the structure information, namely, the density distribution of the projectile and target. The proper target density is employed from electron-scattering experimental data, which is converted to matter densities by unfolding the proton charge density with taking into account the quadrupole deformation. $^{16}$N is divided into two parts for small separation energy of last neutron $E_{1n}$ = 2.491 MeV: $^{15}$N core and a valence neutron part. The harmonic oscillator- (HO-) type function is chosen as the initial function of the core. The single-particle model (SPM) function is chosen as the initial shape of the valence neutron part.

The SPM function is a realistic model to describe the tail structure. The wave function of the valence neutron is calculated by solving the Schödinger equation numerically with Woods-Saxon potential. The SPM takes into account the Coulomb and the centrifugal barrier effects. The main equations of the functions are expressed as

HO type function

$$\rho_c^i(r) = \rho_{c0}^i \times \left(1 + \frac{C-2}{3}\left(\frac{r}{b}\right)^2\right) exp\left(-\left(\frac{r}{b}\right)^2\right), \tag{3}$$

where *i* denotes the proton or neutron, and *C* is the number of protons or neutrons in the core, *b* is the width of the core, and $\rho_{c0}$ is the normalization factor. The same width is used for the proton- and neutron-core density distributions.

Woods-Saxon potential

$$V = \left(-V_0 + V_1(l \cdot s)\frac{r_{ls}^2}{r}\right)\left[1 + exp\left(\frac{r-R_c}{a}\right)\right]^{-1}, \qquad (4)$$

where *a* is the diffuseness parameter, $R_c$ (= $r_0 A^{1/3}$, A is the nuclear mass number) is the radius of the Woods-Saxon potential, $r_{l \cdot s}$ (= 1.1 fm) is the radius for spin orbit potential, and $V_1$ ( = 17 MeV) is the *l·s* strength, taken from [23]. The depth of the potential $V_0$ is adjusted to reproduce the experimental separation energy of the valence neutron.

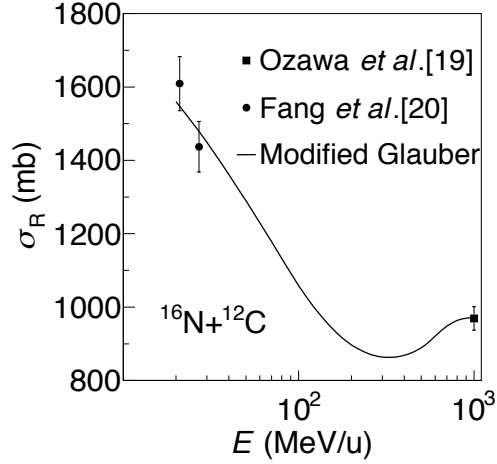

**Fig. 1.** The energy dependent $\sigma_R$ of $^{16}$N on $^{12}$C. The solid curve denotes the theoretical calculation of this work.

Figure 1 shows the energy dependent $\sigma_R$ of $^{16}$N on $^{12}$C target. The best-fit is shown with solid curve; the reduce $\chi^2$ for the best-fit is 0.55, which means a reasonable fit. The width of the HO function is 1.551 ± 0.055fm, and the interaction parameters of the potential are $a_0$ = 0.65 ± 0.13 fm, b = 1.25 ± 0.14 fm. The errors are determined by the method of fit with total $\chi^2$ + 1. The matter radius of $^{16}$N equals 2.385 ± 0.091 fm; it is the low limit of the radii given by previous work through the standard Glauber model, thereby calling for the reanalysis. The difference is easily understandable, because the standard Glauber model underestimate the $\sigma_R$, thus the larger density needed to fill the gap. Figure 2 lists the radius of $^{16}$N.

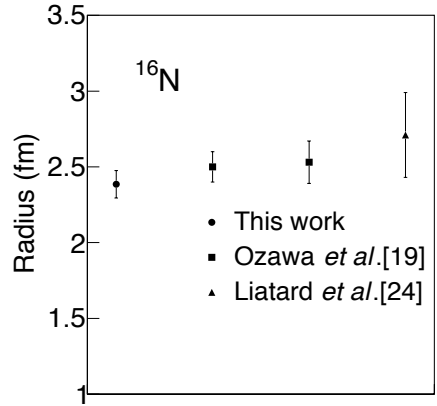

**Fig. 2**. Radius of $^{16}$N. The present result is presented by solid circle. The other results are respresentedby solid squares [19] and solid triangles[24].

## 4 Cross sections of $^{15}$N($n, \gamma$)$^{16}$N

The direct capture for this reaction is dominated by the p→d wave transition to the ground state, p→s wave transition to the first excited state at 0.120 MeV, p→d wave transition to the second excited state at 0.296 MeV, and p→s wave transition to the third excited state at 0.397 MeV of $^{16}$N. The γ-ray transitions are all dominated by the $E1$ multipolarity. The $J_b =2^-$ ground state ($J_b =0^-$ 1st excited state, $J_b =3^-$ 2nd excited state, $J_b =1^-$ 3rd excited state) in $^{16}$N is described as a $j_b = d_{5/2}$ neutron ($j_b = s_{1/2}$ neutron, $j_b = d_{5/2}$ neutron, $j_b = s_{1/2}$ neutron) coupled to the $^{15}$N core, which has an intrinsic spin $I_x = 1/2^-$. Actually, Meissner *et at.* [8] and Huang *et at.* [25] all have made the detailed explainations on the reaction. We will only discuss the errors from the low-lying state of $^{16}$N.

In the calculation, the latest experimental spectroscopic factors by Guo *et al.* are employed. Their values are SF = 0.96 ± 0.09, for the ground state, SF = 0.69 ± 0.09, for the 2$^-$ state, SF = 0.84 ± 0.08, for the 3$^-$ state, and SF = 0.65 ± 0.08, for the 1$^-$ state. The wave function $\psi_b$ of low-lying state in $^{16}$N are calculated numerically in the direct capture theory with $a_0$ =0.65 ± 0.13 fm and b =1.25 ± 0.14 fm obtained in section 3. The parameters for computing the incoming neutron wave function $\psi_c$ are determined with ANC method, as suggested by Huang *et al*. The calculated ANC value Sqrt(SF)b =0.85fm$^{-1/2}$ for the ground state of $^{16}$N, 1.10 fm$^{-1/2}$ for the first excited state, 0.29 fm$^{-1/2}$ for the second excited state, and 1.08 fm$^{-1/2}$ for the third excited state, respectively.

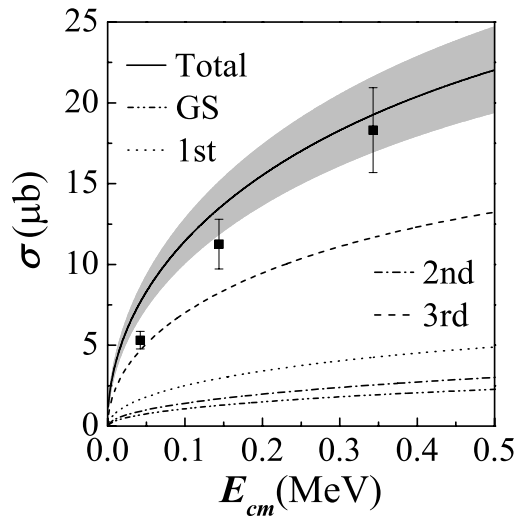

**Fig. 3**. Cross sections of $^{15}$N($n, \gamma$)$^{16}$N. The shaded area indicate the error caused by structure of $^{16}$N, the experimental data are from Ref. [8].

Figure 3 shows the $\sigma_c$ of $^{15}$N + n . The dash dot dot and dash dot lines denote the p→d wave transition to the ground and the second excited states, dot and dash lines denote the p→s transition to the first and third excited states, respectively. The solid line is the summation of four transitions. The shaded area shows the error due to the nuclear properties of $^{16}$N; it is a little higher at 12% comparing to that of $^{7}$Li ($n, \gamma$)$^{8}$Li [12]. There are two main reasons as follows: (a) the same potential parameters are employed to calculate wave functions of $^{16}$N low-lying states without considering the influence of valence neutron to the $^{15}$N core in different state of $^{16}$N; (b) the number of experimental data of $\sigma_R$ existing are not enough to extracted the structure information, the data at intermediate energies ~100 MeV/nucleon are required; (c) the $\sigma_R$ existing contain contributions of excited states, especially the first excited state, the first state in $^{16}$N with life-time of 5.25 μs can reach the reaction target in the transition method. In order to solve the questions noted above, new measurements of the $\sigma_R$ of $^{16}$N are required on stable targets.

## 5 Conclusion

We reanalyzed the reaction cross section of $^{16}$N on $^{12}$C target with modified Glauber model and extracted its structure information. The $^{15}$N + n reaction has been discussed by means of the structure. Although more accurate cross sections are not obtained, the article found problems and the resolution.

The authors would like to thank the support given by the National Natural Science Foundation of China, with Nos. 11447236, 11505002, and 11247001; the Foundation of Anhui University of Science and Technology, with Nos. 11130 and 12608.

References:
[1] C. Rolfs and W. Rodney, Cauldrons in the Cosmos (University of Chicago Press, Chicago, 1988)
[2] L. H. Kawano, W. A. Fowler, R. W. Kavanagh et al., Astrophys. J. 372, 1 (1991)
[3] J. C. Blackmon, A. E. Champagne, J. K. Dickens et al., Phys. Rev. C 54: 383-388 (1996)
[4] S. Burles, K. M. Nollett, J. W. Truran et al, Phys. Rev. Lett. 82, 4176 (1999)
[5] S. E. Woosley, D. H. Hartmann, R. D. Ho_man et al., Astrophys. J. 356, 272 (1990)
[6] M. Forestini, S. Goriely, A. Jorissen et al., Astron. Astrophys. 261, 157 (1992)
[7] A. Jorissen, V. V. Smith, and D. L. Lambert, Astron. Astrophys. 261, 164 (1992)
[8] J. Meissner, H. Schatz, H. Herndl et al., Phys. Rev. C 53, 977 (1996)
[9] W. Bohne, J. Bommer, H. Fuchs et al., Nucl. Phys. A 196, 41 (1972)
[10] D. W. Bardayan et al., Phys. Rev. C 78, 052801(R) (2008)
[11] B. Guo. Z. H. Li, Y. J. Li et al., Phys. Rev. C 89, 012801(R) (2014)
[12] G. W. Fan, M. Fukuda, D. Nishimura et al., Phys. Rev. C 91, 014614 (2015)
[13] I. Tanihata et al., Phys. Rev. Lett. 55, 2676 (1985)
[14] A. Ozawa, T. Suzuki and I. Tanihata, Nucl. Phys. A 693: 32-62 (2001)
[15] R. J. Glauber Lectures in Theoretical Physics. Edited by W. E. Brittin, L. G. Dunham. (New York: Interscience, 1959), p.1: 315
[16] X. Z. Cai et al., Chin. Phys. Lett. 17, 565 (2000)
[17] M. Fukuda et al., Phys. Lett. B 268, 339 (1991).
[18] C. A. Bertulani. Computer Physics Commun., 156: 123 (2003)
[19] A. Ozawa et al., Nucl. Phys. A 691, 599 (2001).
[20] D. Q. Fang et al., Phys. Rev. C 61, 064311 (2000)
[21] M. Takechi, M. Fukuda, M. Mihara et al., Eur. Phys. J. A. 25: 217-219 (2005)
[22] M. Takechi, M. Fukuda, M. Mihara et al., Phys. Rev. C 79, 061601(R) (2009)
[23] A. Bohr and B. R. Mottelson, Nuclear Structure, Vol. I (Benjamin, New York, 1975).
[24] E. Liatard et al., Europhys. Lett. 13, 401 (1990)
[25] J. T. Huang, C. A. Bertulani, and V. Guimartildeaes, At. Data Nucl. Data Tables 96, 824 (2010).